# Boundary-Bulk Interplay in Nonlinear Topological Transport

Deyi Zhuo[1,2], Xiaoda Liu[1,2], Huu-Thong Le[1,2], Annie G. Wang[1], Han Tay[1], Bomin Zhang[1], Ling-Jie Zhou[1], Binghai Yan[1], Chao-Xing Liu[1], and Cui-Zu Chang[1]

[1]Department of Physics, The Pennsylvania State University, University Park, PA 16802, USA

[2]These authors contributed equally: Deyi Zhuo, Xiaoda Liu, and Huu-Thong Le

Corresponding authors: cxl56@psu.edu (C.-X. L.); cxc955@psu.edu (C.-Z. C).

**Abstract: Nonlinear transport has emerged as a powerful approach to probe the quantum geometry of electronic wavefunctions, such as Berry curvature and quantum metric, in topological materials. While nonlinear responses governed by bulk quantum geometry and band topology are well understood, the role of boundary modes (e.g., edge, surface, and hinge states) in nonlinear transport of topological materials remains largely unexplored. In this work, we demonstrate boundary-bulk interplay in nonlinear transport, including second-harmonic Hall and nonreciprocal longitudinal responses, in molecular beam epitaxy-grown magnetic topological insulator heterostructures. We find that the nonlinear transport is maximized when the sample is tuned slightly away from the well-quantized states, including the quantum anomalous Hall and axion insulator states. The sign and amplitude of the nonlinear transport depend on electrode configuration, magnetic order, and carrier type, establishing boundary mode transport as the dominant contributor. These findings, supported by symmetry analysis and nonlinear Landauer-Büttiker formalism, demonstrate that nonlinear transport in topological materials is governed by the interplay between boundary and bulk states. We further derive a universal relation between different lead voltages from electrode geometry symmetry, which allows us to distinguish nonlinear boundary transport from bulk contributions. Our work highlights the critical role of**

**electrodes in nonlinear transport, which is absent in nonlinear optics, and establishes boundary modes as a key origin of the giant nonlinear response in nearly bulk-insulating topological materials. This insight opens new opportunities for engineering nonlinear transport through boundary-bulk interplay in future device applications of topological materials.**

**Main text:** The discovery of topological phases of matter has reshaped our understanding of quantum materials by revealing a profound connection between electronic band structure, symmetry, and topology. Topological materials typically harbor an insulating bulk and symmetry-protected conducting boundary modes (i.e., surface, edge, and hinge states)[1-9]. A prominent example is topological insulators (TIs), in which the gapless nature of surface states is protected by time-reversal ($T$) symmetry[2,3,10-12]. When $T$ symmetry is broken as in magnetic TIs, gapping out surface states leads to a variety of phenomena, such as the quantum anomalous Hall (QAH) effect[13-17] and axion insulator states[5,18-20]. These magnetic TIs can be broadly categorized into ferromagnetic and antiferromagnetic types. Unlike ferromagnetic TIs, antiferromagnetic TIs and magnetic TIs with antiparallel magnetization alignment possess two or more spin sublattices that break inversion ($P$) symmetry and thus induce noncentrosymmetry. The breaking of both $P$ and $T$ symmetry, as well as the combined $PT$ symmetry, induces a finite Berry curvature dipole[21-26] and quantum metric dipole[27-29], which can be probed through second-harmonic electrical transport to uncover the quantum geometry of electronic wavefunctions[30-32]. This emerging framework has motivated increasing interest in nonlinear effects in antiferromagnetic TIs, including recent demonstrations of quantum metric-induced nonlinear Hall and longitudinal responses in exfoliated intrinsic antiferromagnetic TI MnBi$_2$Te$_4$ flakes[27-29]. However, the potential contribution from boundary modes, such as the chiral edge channel of the QAH states in ferromagnetic TIs (Refs.[33-

[35]) and the hinge channel of the axion insulator states in antiferromagnetic TIs (Refs.[6,7,36,37]), has remained elusive, and inspired increasing interest in exploring the boundary-bulk interplay in nonlinear topological transport.

Unlike the mechanical exfoliation of antiferromagnetic crystals, molecular beam epitaxy (MBE) allows for more precise control over material synthesis, providing a stable and reproducible platform for engineering antiferromagnetic ordering through the growth of magnetic TI multilayers with different dopants[5,18-20]. In particular, MBE-grown V-doped $(Bi,Sb)_2Te_3$/$(Bi,Sb)_2Te_3$/Cr-doped $(Bi,Sb)_2Te_3$ sandwiches provide a robust and tunable platform for investigating nonlinear transport, owing to their strong spin-orbit coupling, nontrivial band topology, and the capacity to host well-quantized topological states. These sandwiches can host either QAH or axion insulator states depending on the relative magnetization alignment of the top and bottom surfaces[5,18-20]. An external magnetic field serves as an effective knob to manipulate topological states as well as electrical transport, opening opportunities to realize novel phenomena through variations in the stacking sequence of magnetic TI layers[38,39].

In this work, we employ MBE to grow magnetic TI sandwiches, specifically 3 quintuple layers (QLs) $V_{0.11}(Bi,Sb)_{1.89}Te_3$/4 QL $(Bi,Sb)_2Te_3$/3 QL $Cr_{0.26}(Bi,Sb)_{1.74}Te_3$. Electrical transport measurements reveal that the giant nonlinear transport, including the nonlinear Hall effect and nonreciprocal longitudinal transport, emerges in the nearly bulk-insulating QAH and axion insulator regimes. We find that both the sign and amplitude of the nonlinear Hall and nonreciprocal longitudinal responses depend on electrode configuration, magnetic order, and carrier type. To investigate the underlying mechanism, we conduct symmetry analysis of a six-terminal Hall bar configuration (Fig. 1a) and numerical simulations employing the nonlinear Landauer-Büttiker formalism[40,41]. Our results demonstrate that the observed giant nonlinear responses originate from the local boundary channel, assisted by bulk carriers in the nearly bulk-insulating regime, and thereby display the intricate interplay between bulk and boundary carriers. Our combined experimental and theoretical results underscore the dominant role of boundary modes in governing nonlinear transport in the nearly bulk-insulating regime of topological materials.

All 3 QL $V_{0.11}(Bi,Sb)_{1.89}Te_3$/4 QL $(Bi,Sb)_2Te_3$/3 QL $Cr_{0.26}(Bi,Sb)_{1.74}Te_3$ sandwiches used in this work are grown on heat-treated insulating $SrTiO_3(111)$ substrates in an MBE chamber (Omicron Lab10) with a vacuum better than ~2 × $10^{-10}$ mbar. The Bi/Sb ratio in each layer is optimized to tune the chemical potential of the entire sample near the charge neutral point. Electrical transport measurements are conducted using both a physical property measurement system (PPMS, Quantum Design DynaCool, 1.7 K, 9 T) for $T \geq 1.7$ K and a dilution refrigerator (Leiden Cryogenics, 10 mK, 9 T) for $T < 1.7$ K. The magnetic field is applied perpendicular to the film plane. The same alternating current (AC) is used in first- and second-harmonic electrical transport measurements. Samples S1 to S4 are used in our electrical transport measurements, with Samples S1 and S2 measured in the dilution refrigerator and all samples measured in the PPMS. The main text focuses on Sample S1, and more transport results of Samples S1 to S4 are shown in Extended Data Figs. 1 to 7, 9, and 10, and Figs. S1 to S4.

**QAH and axion insulator states in Magnetic TIs**

We first perform electrical transport measurements to characterize the QAH and axion insulator states in Sample S1 (Fig. 1a). Figure 1b shows the magnetic field $\mu_0H$ dependence of longitudinal conductance $\sigma_{xx}$ and Hall conductance $\sigma_{xy}$ at the charge neutral point $V_g = V_g^0 = 5$ V and $T = 25$ mK. The value of $V_g^0$ for Sample S1 is defined as the gate voltage at which the longitudinal resistance $\rho_{xx}$ reaches its maximum in the axion insulator regime (Extended Data Fig. 1). The sample exhibits the $C = \pm1$ QAH states when the magnetizations of the bottom Cr- and top V-doped (Bi,Sb)$_2$Te$_3$ layer are aligned in parallel. For the $C = \pm 1$ QAH states, the absolute value of the zero magnetic field Hall conductance $|\sigma_{xy}(0)|$ is ~$1.001e^2/h$, concomitant with zero magnetic field longitudinal conductance $\sigma_{xx}(0)$ ~$0.0004e^2/h$ (Fig. 1b). When the magnetizations of the bottom Cr- and top V-doped (Bi,Sb)$_2$Te$_3$ layers are aligned in antiparallel, the sample exhibits two distinct axion insulator states[5,18-20]. These two states are denoted as Axion-I and Axion-II, corresponding to the outward and inward magnetization relative to the surface, respectively. For both Axion-I and Axion-II states, $\sigma_{xx}$ is ~$0.22e^2/h$ and $|\sigma_{xy}|$ ~$0.008e^2/h$ at $\mu_0H = \pm 0.25$ T.

Both the $C = 1$ QAH and the Axion-II states in Sample S1 are further confirmed by the ($V_g$-$V_g^0$) dependence of $\sigma_{xx}$ and $\sigma_{xy}$ at $\mu_0H = 0$ T and $\mu_0H = 0.25$ T after magnetic training, respectively (Figs. 1c and 1d). For the parallel magnetization alignment, $\sigma_{xy}(0)$ shows a quantized plateau at ~$e^2/h$, while $\sigma_{xx}(0)$ vanishes for -15 V ≤ ($V_g$ - $V_g^0$) ≤ 30 V, confirming the well-quantized $C = 1$ QAH state (Fig. 1c). For the antiparallel magnetization alignment, at $\mu_0H = 0.25$ T, $\sigma_{xx}(0.25\text{T})$ exhibits a dip and reaches ~$0.226e^2/h$ at $V_g = V_g^0 = 5$ V, corresponding to the axion insulator state[5,18-20] (Fig. 1d).

**Boundary-bulk interplay in nonlinear transport**

For the linear responses, both Axion-I and Axion-II states exhibit identical $\sigma_{xx}$, while $\sigma_{xy}$ remains close to zero (Fig. 1b). This raises the question of whether the opposite antiparallel magnetization alignments in the Axion-I and Axion-II states manifest in nonlinear Hall and nonreciprocal transport, given that the nonreciprocal response originates from the quadratic term of current $I$ (Ref.[42]). To investigate nonreciprocal transport in the Axion-I and Axion-II states, we measure the second-harmonic longitudinal voltages $V_{xx}^{2\omega}$ at the left and right sides of Sample S1 at different ($V_g$-$V_g^0$) (Fig. 1a), where $V_{xx,L}^{2\omega} = V_6^{2\omega} - V_5^{2\omega}$ and $V_{xx,R}^{2\omega} = V_2^{2\omega} - V_3^{2\omega}$, respectively, with $V_i^{2\omega}$ labelling the second-harmonic voltage at the $i^{th}$ electrode. An AC current $I = 2$ μA is applied to enhance the signal-to-noise ratio and minimize heating effects and other higher-order contributions. We find that the signs of $V_{xx,L}^{2\omega}$ and $V_{xx,R}^{2\omega}$ are opposite to each other and dependent on the carrier type (Figs. 1e, 1f, and Extended Data Fig. 2), which differ significantly from their linear counterparts (Fig. 1b and Extended Data Fig. 1a). In the slightly p-doped region, $V_{xx,L}^{2\omega}$ displays a near-zero plateau in the $C = \pm 1$ QAH regime due to the existence of a chiral edge channel[33-35,43]. $V_{xx,L}^{2\omega}$ exhibits a peak in the Axion-II regime near $\mu_0 H = 0.25$ T, whereas the signal transitions into a dip in the Axion-I regime near $\mu_0 H = -0.25$ T (Fig. 1e). This observation demonstrates that the significant nonreciprocal transport is an intrinsic response for the axion insulator state, depending on outward or inward antiparallel magnetization direction. We note that the polarity of $V_{xx}^{2\omega}$ can be switched by reversing either the *Néel* order orientation or the carrier type. Unlike antiferromagnetic TI MnBi$_2$Te$_4$ flakes[27-29], $V_{xx}^{2\omega}$ in the axion insulator regimes of our magnetic TI sandwiches exhibits a sign reversal between $V_{xx,L}^{2\omega}$ and $V_{xx,R}^{2\omega}$ (Figs. 1e and 1f). This observation demonstrates that $V_{xx}^{2\omega}$ depends on electrode configuration, suggesting the dominance of boundary mode transport within the nearly bulk-insulating axion insulator regimes[5-7,18-20]. This effect cannot be attributed to scattering, as the presence of random magnetic domains near the coercive field[16,44,45] would induce a zero net nonreciprocal transport. Moreover, we exclude the possibility that the second-harmonic voltages observed in our samples arise from gate oscillations[46,47] (Extended Data Fig. 3).

To disentangle the nonlinear boundary transport, we decompose $V_{xx}^{2\omega}$ into symmetrized and antisymmetrized contributions, defined as $V_{xx,S}^{2\omega} = \frac{(V_{xx,R}^{2\omega}+V_{xx,L}^{2\omega})}{2}$ and $V_{xx,A}^{2\omega} = \frac{(V_{xx,R}^{2\omega}-V_{xx,L}^{2\omega})}{2}$, respectively. Here $V_{xx,S}^{2\omega}$ corresponds to the electrode-independent component of $V_{xx}^{2\omega}$, whereas $V_{xx,A}^{2\omega}$ represents the difference between $V_{xx}^{2\omega}$ at the left and right sides and is associated with nonlinear boundary transport. The magnitude of the nonreciprocal coefficient $\gamma$ of the nonlinear boundary transport can be estimated using $\gamma = \sqrt{2}V_{xx,A}^{2\omega}/(R_0 \cdot i^2)$, where $i$ is the root-mean-square of the applied current density[28,42]. The maximum value of $\gamma$ reaches $\sim 1.22 \times 10^{-9}$ m$^2$ A$^{-1}$ at $(V_g\text{-}V_g^0)$ = -5 V in Sample S1, which is an order of magnitude larger than those in exfoliated MnBi$_2$Te$_4$ (Ref.[28]) and WTe$_2$ (Ref.[48]) flakes and four orders of magnitude larger than those reported in heavy metal/ferromagnetic metal heterostructures[49,50].

Figure 2 shows the $\mu_0 H$ dependence of $V_{xx,S}^{2\omega}$, $V_{xx,A}^{2\omega}$ and $V_{yx}^{2\omega}$ (i.e., $V_{yx}^{2\omega} = V_2^{2\omega} - V_6^{2\omega}$) at $T$ = 25 mK over -200 V $\leq (V_g\text{-}V_g^0) \leq$ 200 V. Both the magnitude and sign of $V_{xx,A}^{2\omega}$, $V_{xx,S}^{2\omega}$ and $V_{yx}^{2\omega}$ can be tuned by varying $V_g$. The pronounced contribution of $V_{xx,A}^{2\omega}$ relative to $V_{xx,S}^{2\omega}$ highlights the dominance of nonlinear boundary transport in the total nonlinear response. At $(V_g\text{-}V_g^0)$ = 0 V, $V_{xx,S}^{2\omega}$, $V_{xx,A}^{2\omega}$ and $V_{yx}^{2\omega}$ vanish in both QAH and axion insulator states (Figs. 2c and 2h). This observation indicates that the nonlinear boundary transport is a property associated with chemical potential $E_F$ and requires assistance from bulk carriers[33-35,43]. For the dominant nonlinear boundary contribution in the doped $C$ = 1 QAH regime, $V_{xx,A}^{2\omega}(0)$ increases from -11.8 μV at $(V_g\text{-}V_g^0)$ = -15 V to 0.8 μV at $(V_g\text{-}V_g^0)$ = 100 V (Figs. 2a to 2e), and the value of $V_{yx}^{2\omega}(0)$ correspondingly increases from -11.2 μV at $(V_g\text{-}V_g^0)$ = -15 V to 1.9 μV at $(V_g\text{-}V_g^0)$ = 100 V (Figs. 2f to 2j). In the doped Axion-II regime, the value of $V_{xx,A}^{2\omega}(0.25\text{T})$ decreases from -36.3 μV at $(V_g\text{-}V_g^0)$ = -15 V to a minimum of -101.1 μV at $(V_g\text{-}V_g^0)$ = -5 V to a maximum of 70.1 μV at $(V_g\text{-}V_g^0)$ = 10 V finally to 16.4 μV at $(V_g\text{-}V_g^0)$ = 100 V (Figs. 2a to 2e). The corresponding values of $V_{yx}^{2\omega}(0.25\text{T})$ are -70.8 μV, -130.8 μV, 83.9 μV, and 20.7 μV at $(V_g\text{-}V_g^0)$ = -15 V, -5 V, 10 V, and 100 V, respectively (Figs. 2f to 2j). We note that the nonlinear Hall voltage $V_{yx}^{2\omega}$ is comparable with $V_{xx,R}^{2\omega}$ as a function of $\mu_0 H$ across different $V_g$ values, which is constrained by the symmetry of the electrode geometry (Methods). Specifically, for slightly p- and n-doped regions, the $V_{yx}^{2\omega}$ curves almost coincide with the $V_{xx,R}^{2\omega}$ curves (Figs. 2g and 2i). $V_{xx,L}^{2\omega}$, $V_{xx,R}^{2\omega}$ and $V_{yx}^{2\omega}$ are frequency-independent and scale quadratically with the applied AC current (Extended Data Figs. 4 and 5). The nonlinear responses vanish near the Curie temperature $T_C \sim 20$ K of Sample S1(Extended Data Fig. 6).

A spike feature is observed in both $V_{xx,A}^{2\omega}$ and $V_{yx}^{2\omega}$ during the quantum phase transition between the QAH and axion insulator states (Fig. 2). This behavior resembles the nonzero nonlinear response observed during the quantum Hall plateau phase transitions and can be attributed to additional dissipation from the surface states during magnetization reversal[43]. When $\mu_0 H$ sweeps near the coercive field of the bottom Cr-doped $(Bi,Sb)_2Te_3$ layer, the presence of random magnetic domains and associated scattering introduces additional dissipative channels, resulting in a sudden spike in $|V_{xx,A}^{2\omega}|$ and $|V_{yx}^{2\omega}|$. Therefore, $V_{xx,A}^{2\omega}$ and $V_{yx}^{2\omega}$ retain their signs until the sample fully transitions to the axion insulator state. Furthermore, the extrema in the $\mu_0 H$ dependence of $V_{xx,A}^{2\omega}$ and $V_{yx}^{2\omega}$ consistently emerge in the Axion-I and Axion-II regimes. In the antiparallel magnetization configuration, the values of $|V_{xx,A}^{2\omega}|$ and $|V_{yx}^{2\omega}|$ are larger in the slightly doped region than in the highly doped region, implying the critical role of in-band boundary states[7] buried in the metallic surface states for nonlinear response. This observation demonstrates that the giant nonlinear signals are not purely induced by bulk carriers but instead result from the interplay between boundary and bulk carriers.

For the electrode-independent nonlinear longitudinal response, a significant $V_{xx,S}^{2\omega}$ emerges in the Axion-I and Axion-II regimes within the slightly doped regions (Figs. 2b and 2d). The sign reversal of $V_{xx,S}^{2\omega}$ between the two axion insulator states with opposite *Néel* order is consistent with the quantum metric-induced nonlinear bulk transport in exfoliated $MnBi_2Te_4$ flakes[28] and stands in stark contrast to Berry curvature dipole-induced responses, which allow nonlinear Hall signals but not longitudinal signals[22,23]. As shown in Figs. 2a to 2e, the nonlinear boundary transport ($V^{2\omega}_{xx,A}$) dominates over the electrode-independent nonlinear transport ($V^{2\omega}_{xx,S}$) at different ($V_g$-$V_g^0$), indicating the robustness of boundary modes in the axion insulator regime.

Figure 3 and Extended Data Fig. 7 show the nonlinear responses as a function of ($V_g$-$V_g^0$) for the $C = 1$ QAH and Axion-II states at $T = 25$ mK. Both $V^{2\omega}_{xx,A}$ and $V^{2\omega}_{yx}$ exhibit distinct behaviors in the $C = 1$ QAH and Axion-II states. For -5 V $\leq$ ($V_g$-$V_g^0$) $\leq$ 10 V, where $E_F$ is tuned within the magnetic exchange gap, both signals exhibit a near-zero plateau in the $C = 1$ QAH regime around $V_g = V_g^0$ due to the persistence of the chiral edge channel. In contrast, in the Axion-II state, both signals display a rapid sign reversal over -5 V $\leq$ ($V_g$-$V_g^0$) $\leq$ 10 V. Upon tuning the sample away from $V_g = V_g^0$, the magnitudes of $|V^{2\omega}_{xx,A}|$ and $|V^{2\omega}_{yx}|$ in the $C = 1$ QAH state first increase and then decrease as bulk transport becomes dominant. Furthermore, both $|V^{2\omega}_{xx,A}|$ and $|V^{2\omega}_{yx}|$ in the $p$-doped region [i.e., ($V_g$-$V_g^0$) < 0 V] exhibit a rapidly varying dip behavior, in sharp contrast to the slow variation in the $n$-doped region [ i.e., ($V_g$-$V_g^0$) > 0 V]. This observed asymmetric behavior is consistent with the asymmetric gate dependence in the linear conductance in Fig. 1c and can be attributed to the band structure of the magnetic TI, where the magnetic exchange gap lies closer to the bulk valence band maximum but further from the bulk conduction band minimum, consistent with the prior studies[51-54]. For the Axion-II state, both $|V^{2\omega}_{xx,A}|$ and $|V^{2\omega}_{yx}|$ exhibit a peak or dip behavior for both $n$- and $p$-doped regions, reflecting the evolution of in-band boundary contributions into nonchiral top and bottom surface contributions as $E_F$ is tuned away from $V_g = V_g^0$.

**Symmetry constraints on nonlinear transport**

In the theoretical framework, the nonlinear response of photocurrents (e.g., bulk photovoltaic effect) and bulk transport are analyzed using the second-order bulk conductivity $\sigma_{\alpha\beta\gamma}$, defined by $j_\alpha = \sum_{\beta\gamma} \sigma^{(2)}_{\alpha\beta\gamma} E_\beta E_\gamma$, where $j_\alpha$ is the current density, $E_{\beta(\gamma)}$ is the external electric field and $\alpha, \beta, \gamma = x, y, z$. The form of the second-order conductivity tensor $\sigma^{(2)}_{\alpha\beta\gamma}$ is strongly constrained by the symmetry of the microscopic Hamiltonian. For example, $\sigma^{(2)}_{\alpha\beta\gamma}$ must vanish in three-dimensional centrosymmetric crystals with $P$ symmetry[55] or two-dimensional crystals with two-fold rotation $C^{2z}$ along the $z$ axis with $\alpha, \beta, \gamma$ only chosen to be $x, y$ for $\sigma^{(2)}_{\alpha\beta\gamma}$. However, this symmetry analysis based on the bulk conductivity $\sigma^{(2)}_{\alpha\beta\gamma}$ does not involve the electrode configuration of transport measurements and cannot explain the prominent electrode configuration-dependent nonlinear response observed in our experiments, which is intimately related to boundary modes.

To develop a systematic symmetry analysis for the experimental setup in Fig. 1a, we consider the second-order conductance $G^{(2)}_{ijk}$, instead of conductivity $\sigma^{(2)}_{\alpha\beta\gamma}$, for nonlinear topological transport, with $G^{(2)}_{ijk}$ defined by $I_i = \sum_{jk} G^{(2)}_{ijk} V_j V_k$, where $i, j, k = 1, \dots, 6$ label different electrodes in Fig. 1a instead of spatial coordinates (e.g. $x, y$), and $I_i$ and $V_j$ denote the current and voltage in the $i^{th}$ and $j^{th}$ electrodes, respectively. The symmetry property of $G^{(2)}_{ijk}$ not only depends on the crystal symmetry of the microscopic Hamiltonian but also is constrained by the symmetry of the sample geometry and electrode configurations (referred to as geometry symmetry below). We consider the standard Hall bar with six terminals (Fig. 1a) for magnetic TI models (Methods), in which both the crystal symmetry and geometry symmetry involve a two-fold rotation $C^{2z}$. The $C^{2z}$ symmetry interchanges different electrodes ($1 \leftrightarrow 4, 2 \leftrightarrow 5, 3 \leftrightarrow 6$ in Fig. 1a), thus serving as a permutation matrix, denoted as $\mathcal{C}^{2z}$, for the current and voltage vectors, $\boldsymbol{I} = (I_1, I_2, \dots, I_6)$ and $V = (V_1, V_2, ..., V_6)$ (Methods). From the Neumann's principle[56,57], the $C^{2z}$ symmetry requires the constraint $G_{ijk}^{(2)} = \sum_{i'j'k'} C_{ii'}^{2z} C_{jj'}^{2z} C_{kk'}^{2z} G_{i'j'k'}^{(2)}$. We consider an AC current $I^\omega = (I, 0, 0, -I, 0, 0)$ from electrode 1 to 4 with the frequency $\omega$ and analyze the second-harmonic voltages $V^{2\omega}$ at different electrodes to extract $V_{xx,L}^{2\omega}$, $V_{xx,R}^{2\omega}$ and $V_{yx}^{2\omega}$.

Using the form of the driving current $I^\omega$, which satisfies $C^{2z} I^\omega = -I^\omega$, we derive the constraints $V_1^{2\omega} = V_4^{2\omega}, V_2^{2\omega} = V_5^{2\omega}, V_3^{2\omega} = V_6^{2\omega}$ from the $C^{2z}$ symmetry (Methods and Supplementary Information), which give rise to $V_{xx,L}^{2\omega} = -V_{xx,R}^{2\omega} = -V_{yx}^{2\omega}$. Therefore, based on the symmetry analysis that takes into account the electrode configuration, we can explain two key features observed in our experiments: (*i*) The opposite signs between $V_{xx,R}^{2\omega}$ and $V_{xx,L}^{2\omega}$; (*ii*) The similar qualitative behaviors as a function of gate voltages and similar amplitude of $V_{xx,R}^{2\omega}$ and $V_{yx}^{2\omega}$, both originating from the approximate $C^{2z}$ symmetry of our Hall bar devices. We note that our symmetry analysis is general for boundary-dominant nonlinear transport with bulk suppressed by $C^{2z}$ symmetry and does not rely on the microscopic model or whether the transport is diffusive or ballistic.

**Nonlinear Landauer-Büttiker formalism**

Next, we consider a two-surface-state model that consists of two Dirac surface states at the top and bottom surfaces with magnetic exchange gaps and numerically simulate the nonlinear transport behavior of this model using the nonlinear Landauer-Büttiker formalism[40,41] that takes into account the influence of the electrode configuration (Methods and Supplementary Information). Figure 4 and Extended Data Fig. 8 present the simulated linear and nonlinear responses as a function of carrier density for parallel and antiparallel magnetization configurations which demonstrate $V_{xx,L}^{2\omega} = -V_{xx,R}^{2\omega} = -V_{yx}^{2\omega}$, in good agreement with our symmetry analysis. In addition, our numerical simulations reveal several qualitative features beyond symmetry analysis: (*i*) Our simulations show sign changes of $V_{xx,R}^{2\omega}$ and $V_{yx}^{2\omega}$ upon the reversal of magnetization in both parallel and antiparallel magnetization alignments. This behavior is explained by acting the operation $M_y$ on the effective model for the magnetic TI sandwich (Supplementary Information); (*ii*) The sign reversal of nonlinear voltages between *n*- and *p*-doped regions, as well as the identical behaviors of $V_{xx,R}^{2\omega}$ and $V_{yx}^{2\omega}$ in both parallel and antiparallel magnetization alignments are observed, which can qualitatively explain the ($V_g$-$V_g^0$) dependence of $V_{xx,R}^{2\omega}$ and $V_{yx}^{2\omega}$ for both the *C* = 1 QAH and Axion-II regimes (Fig. 3). As the asymmetric behavior in Fig. 3 originates from bulk carriers, which is not captured in the two-surface-state model, we include additional parabolic valence bands in our numerical simulations and find the enhanced nonlinear voltage in the *p*-doped region compared to the *n*-doped region (Fig. S10); (*iii*) $V_{xx,R}^{2\omega}$ and $V_{yx}^{2\omega}$ in the axion insulator regime are much larger than those in the QAH regime, due to a larger transmission probability derivative, e.g. $\partial_E T_{ij}$ and $\partial_{V_k} T_{ij}$, in the axion insulator regime (Fig. S11). We further confirm that the hinge current due to the in-band boundary states[7] flowing at the top and bottom surfaces in the axion insulator regime plays a substantial role (Supplementary Information).

**Discussion and outlook**

Finally, we discuss the unique properties of nonlinear boundary transport in nearly bulk-insulating magnetic TI sandwiches. First, the nonlinear boundary transport is more prominent than the nonlinear bulk response associated with quantum geometry and bulk band topology in this regime. The nonlinear transport from residual bulk states may be suppressed due to *P* or $C^{2z}$ symmetry[58], therefore, a nearly bulk-insulating system is necessary for the observation of giant boundary dominant nonlinear transport in topological materials. Second, the nonlinear boundary transport mediated by the axion insulator state exhibits distinct signatures for opposite *Néel* orders and remains robust against external magnetic field perturbations. This transport property shows reproducibility and consistency across multiple samples (Extended Data Figs. 9 and 10, and Figs. S3 and S4). Third, the nonlinear boundary transport can be effectively tuned through external $\mu_0 H$ and/or $V_g$. Fourth, our symmetry analysis reveals that even though nonlinear bulk conductivity vanishes due to the $C^{2z}$ symmetry, the nonlinear boundary modes can still give rise to second-harmonic responses in nonlinear transport. This behavior contrasts with linear topological transport, in which the boundary-bulk correspondence implies that the linear Hall conductivity from the bulk calculation should coincide with the boundary transport from the chiral edge states in the thermodynamic limit based on the Landauer-Büttiker formalism[59,60]. Our combined experimental and theoretical findings highlight the importance of the interplay between boundary and bulk carriers in nonlinear topological transport.

To summarize, we demonstrate that a giant nonlinear boundary transport dominates in our slightly doped QAH and axion insulator samples, where a nearly insulating bulk is needed, consistent with our theoretical analysis and calculations. The sign of the nonlinear topological responses is determined by the direction of the *Néel* order or ferromagnetic order, the electrode positions at the edge, and the $E_F$. The opposite sign of nonlinear voltages for opposite *Néel* order directions provides an electrical transport signal to measure antiferromagnetic *Néel* order. The magnitude of nonreciprocal coefficient $\gamma$ is much larger than those in prior works on topological materials[28,48-50]. The electrode geometry dependence serves as a useful tool to distinguish nonlinear boundary transport from bulk transport. Our research advances the understanding of the interplay between nonlinear boundary and bulk transport in topological materials, facilitating the development of next-generation electronic devices, including high-performance wireless rectifiers[24,61-63].

**Methods**

**MBE growth**

All 3 QL $V_{0.11}(Bi,Sb)_{1.89}Te_3$/4 QL $(Bi,Sb)_2Te_3$/3 QL $Cr_{0.26}(Bi,Sb)_{1.74}Te_3$ sandwiches used in this work are grown in a commercial MBE chamber (Omicron Lab10) with a base vacuum better than ~$2 \times 10^{-10}$ mbar. The heat-treated ~0.5 mm thick $SrTiO_3$(111) substrates are first outgassed at ~600 °C for 1 hour before the MBE growth. High-purity Bi (99.9999%), Sb (99.9999%), Cr (99.999%), V (99.995%), and Te (99.9999%) are evaporated from Knudsen effusion cells. During MBE growth, the substrate is maintained at ~230 °C. The Te/(Bi + Sb + Cr/V) flux ratio is set to be greater than ~10 to prevent Te vacancies in the films. The Bi/Sb ratio in each layer is optimized to tune the chemical potential of the entire magnetic TI sandwich heterostructure near the charge neutral point. The growth rate of both undoped and magnetically doped TI layer is ~0.2 QL per minute. No capping layer is involved in our ex situ electrical transport measurements.

**Electrical transport measurements**

All magnetic TI sandwich heterostructures grown on 2 mm × 10 mm insulating $SrTiO_3$(111) substrates are scratched into a Hall bar geometry using a computer-controlled probe station. The effective area of the Hall bar is ~1 mm × 0.5 mm. The electrical ohmic contacts are made by pressing indium dots onto the films. The bottom gate is prepared by flattening the indium dots on the back side of the $SrTiO_3$(111) substrates. Electrical transport measurements are conducted using both a Physical Property Measurement System (Quantum Design DynaCool, 1.7 K, 9 T) for $T \geq$ 1.7 K and a dilution refrigerator (Leiden Cryogenics, 10 mK, 9 T) for $T <$ 1.7 K. The magnetic field is applied perpendicular to the sample plane. The bottom gate voltage $V_g$ is applied using a Keithley 2450 source meter. All magneto-transport results shown in this work are symmetrized or anti-symmetrized as a function of the magnetic field to eliminate the influence of the electrode misalignment.

**Symmetry constraints on second-harmonic voltages in a $C_{2z}$-symmetric system**

We use the following equations to describe the nonlinear transport in our experiment:

$$\sum_j G_{ij}^{(1)} V_j^{2\omega} = -\sum_{jk} G_{ijk}^{(2)} V_j^{\omega} V_k^{\omega} \tag{1}$$

$$\sum_j G_{ij}^{(1)} V_j^{\omega} = I_i^{\omega} \tag{2}$$

up to the second order of the driving current $I_i^{\omega}$. The Neumann's principle requires the symmetry constraints on the conductance as $G_{ij}^{(1)} = \sum_{i'j'} C_{ii'}^{2z} C_{jj'}^{2z} G_{i'j'}^{(1)}$ and $G_{ijk}^{(2)} = \sum_{i'j'k'} C_{ii'}^{2z} C_{jj'}^{2z} C_{kk'}^{2z} G_{i'j'k'}^{(2)}$, where the matrix $C^{2z}$ is given by

$$C^{2z} = \begin{pmatrix} 0 & 0 & 0 & 1 & 0 & 0 \\ 0 & 0 & 0 & 0 & 1 & 0 \\ 0 & 0 & 0 & 0 & 0 & 1 \\ 1 & 0 & 0 & 0 & 0 & 0 \\ 0 & 1 & 0 & 0 & 0 & 0 \\ 0 & 0 & 1 & 0 & 0 & 0 \end{pmatrix} \tag{3}$$

Furthermore, the current configuration used in the six-terminal Hall bar satisfies $\sum_j C_{ij}^{2z} I_j^{\omega} = -I_i^{\omega}$.

With these equations, we find that the second-harmonic voltages satisfy

$$\sum_{jk} G_{ij}^{(1)} (\mathbf{1} - C^{2z})_{jk} V_k^{2\omega} = 0, \quad \forall i \tag{4}$$

which implies that $\sum_j C_{ij}^{2z} V_j^{2\omega} = V_i^{2\omega}$. Therefore, $V_i^{2\omega}$ are perfectly invariant under the transformation of the matrix $C^{2z}$. This constraint leads to $V_1^{2\omega} = V_4^{2\omega}$, $V_2^{2\omega} = V_5^{2\omega}$, $V_3^{2\omega} = V_6^{2\omega}$ or $V_{xx,L}^{2\omega} = -V_{xx,R}^{2\omega} = -V_{yx}^{2\omega}$, indicating that the nonlinear voltages flip sign when measured at the left and right sides of the sample. Moreover, this proves that the second-harmonic longitudinal and Hall voltages must have the same magnitude in a $C^{2z}$-symmetric system. More details are provided in Supplementary Information.

**Theoretical model and nonlinear Landauer-Büttiker formalism**

To model our magnetic TI sandwich structure, we consider the two-surface-states model, described by the Hamiltonian

$$H_{0,t/b} = v_F(k_y\sigma_x - k_x\sigma_y)\tau_z + M_+\sigma_z + M_-\sigma_z\tau_z + (m_0 + Bk^2)\tau_x, \qquad (5)$$

where $\sigma_{x,y,z}$ and $\tau_{x,y,z}$ are the Pauli matrices for spin and layer degrees of freedom, respectively; $v_F$ is the Fermi velocity, $M_\pm = (M_t + M_b)/2$ with $M_{t/b}$ representing the top/bottom surface magnetization, and $m_0 + Bk^2$ is the inter-layer tunneling. We consider both the parallel ($M_t M_b > 0$) and antiparallel magnetization alignment configurations ($M_t M_b < 0$). Illustration of band structure and wave functions are shown in Fig. S8. We regularize this model into a tight-binding model on a square lattice with the size $L_x = 100, L_y = 20$ for the study of nonlinear transport in magnetic TI sandwiches. The parameters chosen are $v_F = 2, m_0 = 0.1, B = -1, M_t = \pm 1.2, M_b = \pm 0.6$ along with the lattice constant $a = 1$. The Hamiltonian (5) serves as the low-energy effective theory of our system and respect two-fold rotation symmetry, i.e., a quasi-symmetry of the realistic system in the context of the approximate symmetry in the $\boldsymbol{k} \cdot \boldsymbol{p}$ theory[64]. This is a reasonable approximation as the pronounced nonlinear topological transport signals are observed close to either QAH or axion insulator states in our experiments, for which the higher-order terms that break two-fold rotation symmetry are irrelevant.

To simulate the nonlinear transport using the Landauer-Büttiker formalism[40,41], we consider the electrode configuration in Fig. S5, and the current $I_i$ in the $i^{th}$ electrode can be related to the voltages by $I_i = \sum_j G_{ij}^{(1)} V_j + \sum_j G_{ijk}^{(2)} V_j V_k$ up to the second order in voltages, where the conductance matrices are given by

$$G_{ij}^{(1)} = \frac{e^2}{h}\left[-T_{ij} + \delta_{ij}\sum_{j'} T_{ij'}\right] \quad (6)$$

$$G_{ijk}^{(2)} = -g_{ijk}^{(2)} + \delta_{ij}\sum_{j'} g_{ij'k}^{(2)} \quad (7)$$

$$g_{ijk}^{(2)} = \frac{e^3}{h}\left[\frac{1}{2}(\delta_{ik} + \delta_{jk})\partial_E T_{ij} + \partial_{eV_k} T_{ij}\right]. \quad (8)$$

We numerically evaluate the transmission probability $T_{ij}$ in the six-terminal setup in Fig. 1a using the KWANT package for quantum transport[65], where $i, j = 1, \ldots, 6$ label six electrodes. Using the calculated $G_{ij}^{(1)}, G_{ijk}^{(2)}$ and the current configuration $\boldsymbol{I}^{\omega} = (I, 0, 0, -I, 0, 0)$, we choose $I = 1$ and calculate the second-harmonic voltage vector $\boldsymbol{V}^{2\omega}$, from which we can extract $V_{xx,L}^{2\omega}$, $V_{xx,R}^{2\omega}$ and $V_{yx}^{2\omega}$. From Eqs. 6 to 8, the first-order conductance $G_{ij}^{(1)}$ reflects the scattering of electrons among different electrodes and is proportional to the transmission matrix $T_{ij}$. On the other hand, the second-order conductance $G_{ijk}^{(2)}$ arises from the derivative of transmission matrix $T_{ij}$ with respect to electron energy or external electrode voltages, namely $\partial_E T_{ij}$ and $\partial_{V_k} T_{ij}$. In the expression of $g_{ijk}^{(2)}$ above, the first term is proportional to $\partial_E T_{ij}$ purely stems from the redistribution of charges in the electrodes, while the second term is proportional to $\partial_{eV_k} T_{ij}$ and stems from the self-consistent electrostatic potential $eU(r)$ due to the redistribution of charges inside the system. The latter is crucial for gauge invariance under a global energy shift in all gate voltages[40].

We study the nonlinear transport in the low-frequency driving current, which can be derived as $\sum_j G_{ij}^{(1)} V_j^{2\omega} = -\sum_{jk} G_{ijk}^{(2)} V_j^{\omega} V_k^{\omega}$ and $\sum_j G_{ij}^{(1)} V_j^{\omega} = I_i^{\omega}$ from the current-voltage relations given above (Supplementary Information). Given the first- and second-order conductance, inverting these equations would allow us to calculate the voltages from the current configurations, which directly gives the voltages measured in experiments. However, the matrix $G^{(1)}$ is not invertible since its rank is not full (Supplementary Information). We can define $\tilde{I} = [I_1, ..., I_{N-1}]$, and $\tilde{V} = [V_1, ..., V_{N-1}] - V_N$, with $N$ is the number of electrodes, to derive the relation

$$\tilde{I}_i^\omega = \tilde{G}_{ij}^{(1)} \tilde{V}_j^\omega \quad \& \Leftrightarrow \quad \tilde{V}_i^\omega = \tilde{R}_{ij}^{(1)} \tilde{I}_j^\omega \tag{9}$$

$$\tilde{G}_{ij}^{(1)} \tilde{V}_j^{2\omega} = -\tilde{G}_{ijk}^{(2)} \tilde{V}_j^\omega \tilde{V}_k^\omega \quad \& \Leftrightarrow \quad \tilde{V}_i^{2\omega} = -\tilde{R}_{ii'}^{(1)} \tilde{G}_{i'j'k'}^{(2)} \tilde{R}_j^{(1)} \tilde{I}_{jj'}^\omega \tilde{R}_k^{(1)} \tilde{I}_{kk'}^\omega \tag{10}$$

where $\tilde{G}^{(1)}$ is the matrix $G^{(1)}$ with the last row and column removed, $\tilde{R}^{(1)} = [\tilde{G}^{(1)}]^{-1}$, and $\tilde{G}^{(2)}$ is $G^{(2)}$ with the $N^{th}$ values of second and third dimensions removed. Knowing the second-harmonic voltage $\tilde{V}_i^{2\omega}$, we would be able to compare the numerical simulation directly with the transport experiment measurements.

**Acknowledgments:** We thank A. Akhmerov, L. Fu, J. E. Moore, D. Xiao, S. Xu, and Y. Zhao for helpful discussions. This project is primarily supported by the ARO award (W911NF2210159), including MBE growth and dilution transport measurements. Sample characterization and theoretical calculations are supported by the Seed project of the Penn State MRSEC for Nanoscale Science (DMR-2011839). PPMS measurements are supported by the NSF grant (DMR-2241327) and the ONR Award (N000142412133). CZC acknowledges the support from the Gordon and Betty Moore Foundation's EPiQS Initiative (GBMF9063 to C. -Z. C).

**Author contributions:** CZC conceived and designed the experiment. DZ, HT, AGW, and CZC performed MBE growth and PPMS measurements. XL, DZ, BZ, LJZ, and CZC performed dilution transport measurements. HTL, BY, CXL provided theoretical support. DZ, HTL, CXL, and CZC analyzed the data and wrote the manuscript with input from all authors.

**Competing interests:** The authors declare no competing financial interests.

**Data availability:** The data that support the findings of this article are openly available [66].

**Figures and figure captions**

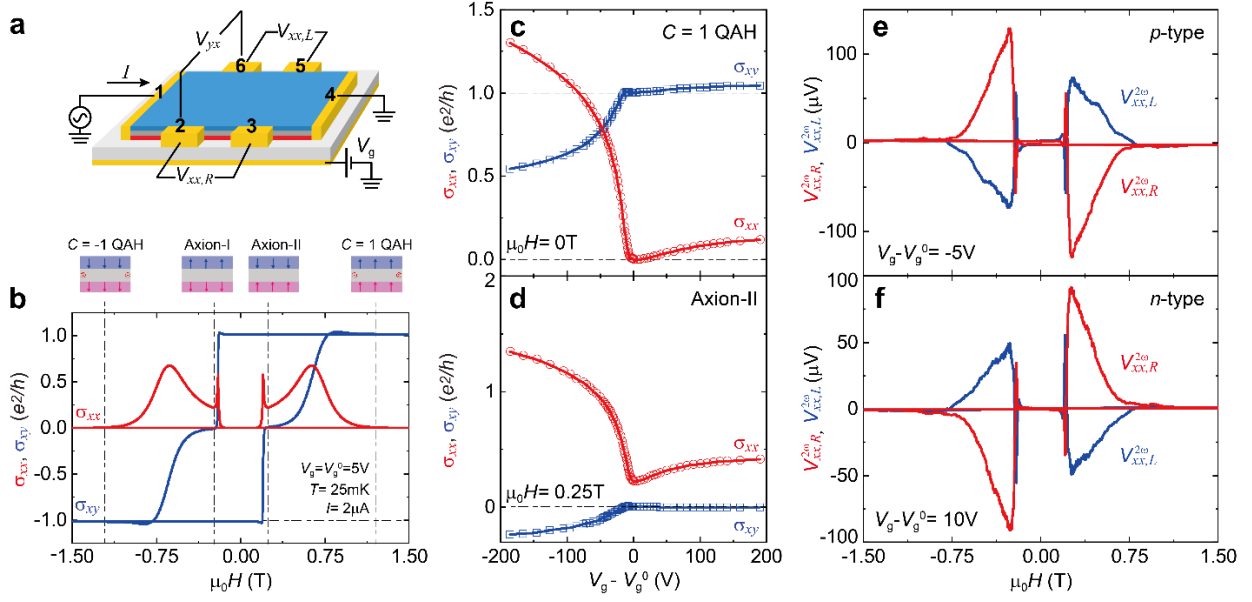

**Fig. 1| Boundary-bulk interplay in nonreciprocal longitudinal responses. a,** Schematic of the magnetic TI sandwich Hall bar device. The 3 QL V-doped $(Bi,Sb)_2Te_3$, 4 QL $(Bi,Sb)_2Te_3$, 3 QL Cr-doped $(Bi,Sb)_2Te_3$ layers are shown in blue, grey, and red, respectively. $V_{xx,R}(V_{xx,L})$ denotes the longitudinal voltage measured at the right (left) edge. **b,** $\mu_0H$-dependent $\sigma_{xx}$ (red) and $\sigma_{xy}$ (blue) measured at $V_g = V_g^0 = 5$ V and $T = 25$ mK. The $C = \pm 1$ QAH and the two types of axion insulator states with different magnetization alignments (Axion-I and Axion-II) are indicated by the dashed lines. **c,** $(V_g-V_g^0)$-dependent $\sigma_{xx}(0)$ (red) and $\sigma_{xy}(0)$ (blue) for the $C = 1$ QAH state measured at $\mu_0H = 0$ T and $T = 25$ mK. **d,** $(V_g-V_g^0)$-dependent $\sigma_{xx}(0.25T)$ (red) and $\sigma_{xy}(0.25T)$ (blue) for the Axion-II state measured at $\mu_0H = 0.25$ T and $T = 25$ mK. **e, f,** $\mu_0H$-dependent $V_{xx,R}^{2\omega}$ (red) and $V_{xx,L}^{2\omega}$ (blue) measured at $(V_g-V_g^0) = -5$ V (**e**) and 10 V (**f**) and $T = 25$ mK.

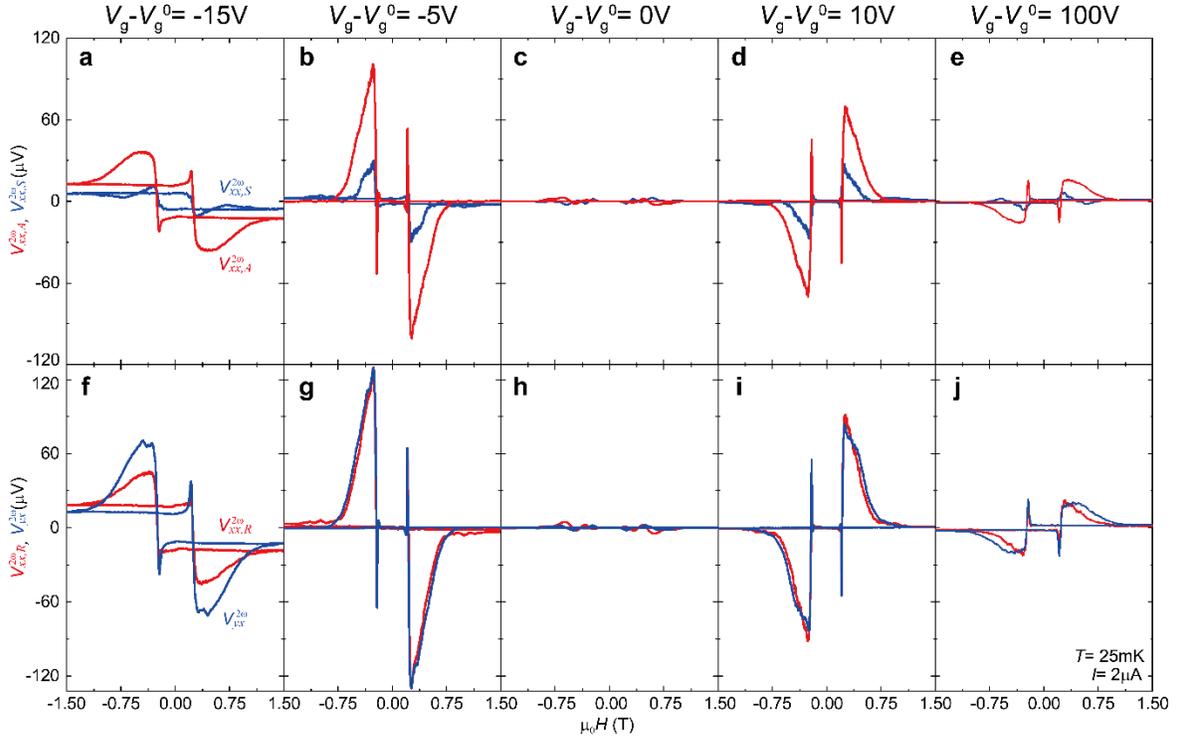

**Fig. 2| Nonlinear transport from boundary-bulk interplay measured across different gate voltages. a-e,** $\mu_0H$-dependent $V^{2\omega}_{xx,A}$ (red) and $V^{2\omega}_{xx,S}$ (blue) measured at $(V_g-V_g^0)$ = -15 V (**a**), -5V (**b**), 0 V (**c**), 10 V (**d**), and 100 V (**e**). **f-j,** $\mu_0H$-dependent $V^{2\omega}_{xx,R}$ (red) and $V^{2\omega}_{yx}$ (blue) measured at $(V_g-V_g^0)$ = -15 V (**f**), -5 V (**g**), 0 V (**h**), 10 V (**i**), and 100 V (**j**). All measurements are performed at $T$ = 25 mK.

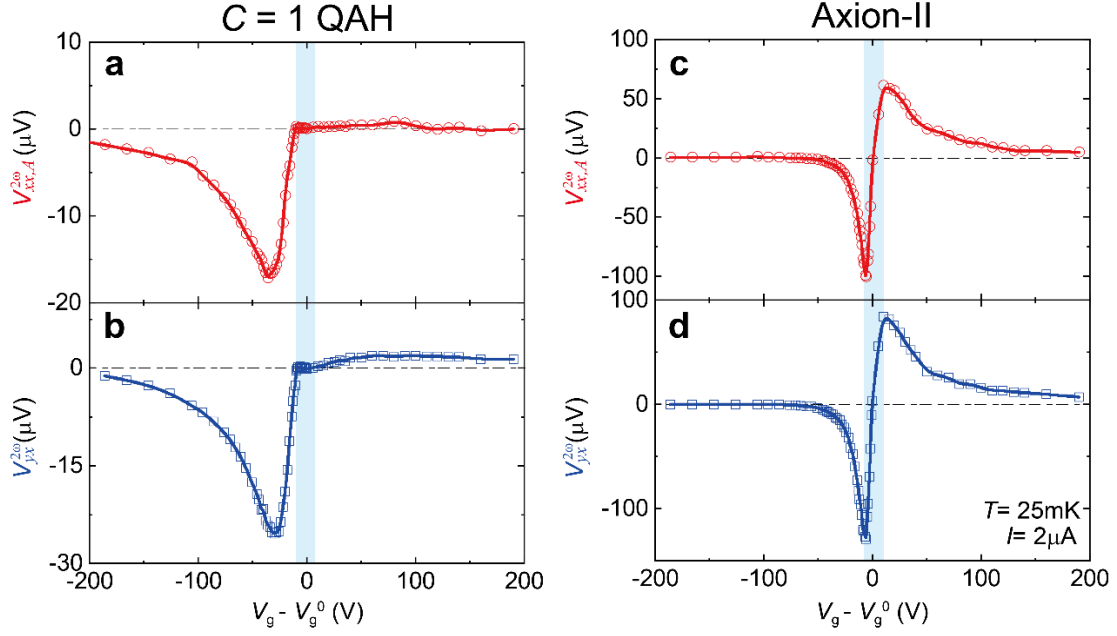

**Fig. 3| Gate-tunable nonlinear transport from boundary-bulk interplay. a, b,** ($V_g$-$V_g^0$)-dependent $V_{xx,A}^{2\omega}$ (**a**) and $V_{yx}^{2\omega}$ (**b**) for the $C = 1$ QAH state. **c, d**, ($V_g$-$V_g^0$)-dependent $V_{xx,A}^{2\omega}$ (**c**) and $V_{yx}^{2\omega}$ (**d**) for the Axion-II state.

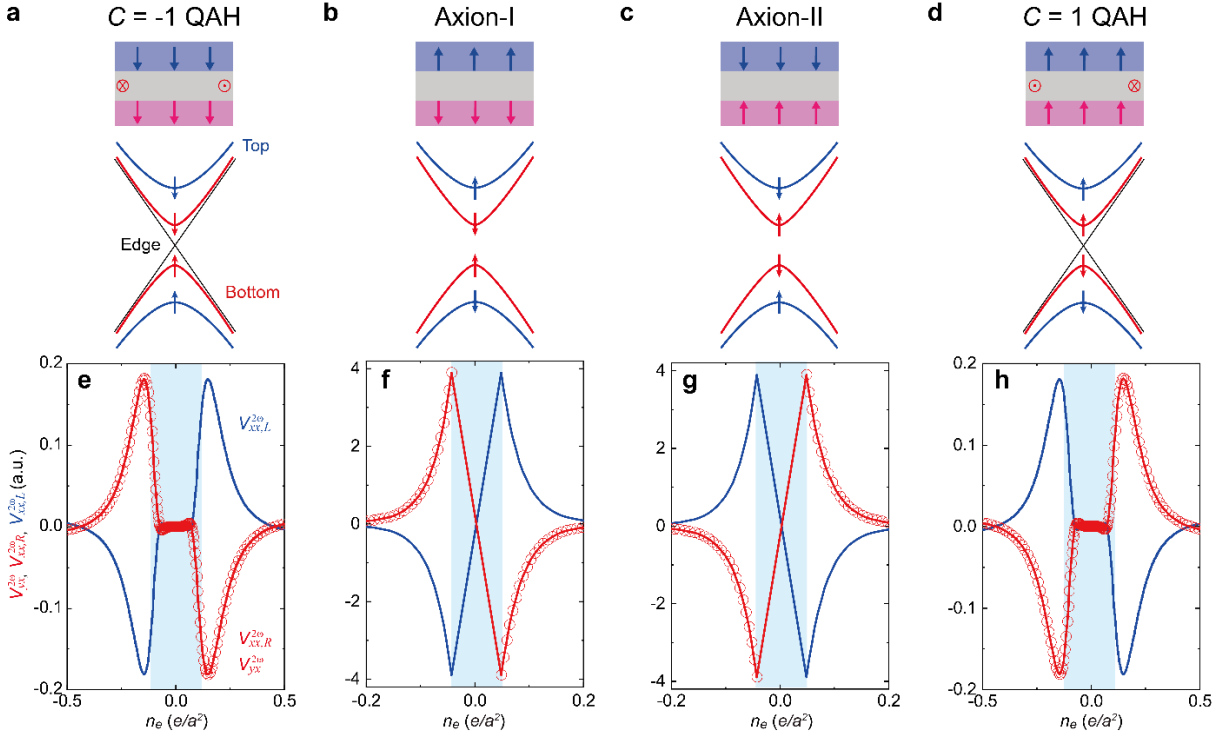

**Fig. 4| Calculated second-order voltages as a function of carrier density for different quantum states. a-d,** Schematics of the magnetic TI sandwich with different magnetization alignments and the corresponding surface energy band dispersion based on the two-surface-state model. Colored arrows denote the spin polarization at the two surfaces, with red (blue) representing the top (bottom) surface. **e-h,** The calculated second-harmonic voltages $V_{xx,L}^{2\omega}$ (blue line), $V_{xx,R}^{2\omega}$ (red line), and $V_{yx}^{2\omega}$ (red circles) as a function of $n_e$ for the quantum states shown in (**a-d**), where we find $V_{xx,R}^{2\omega} = V_{yx}^{2\omega}$. The magnetic exchange gap regime is highlighted by the blue-shaded region.

**Extended Data Figures:**

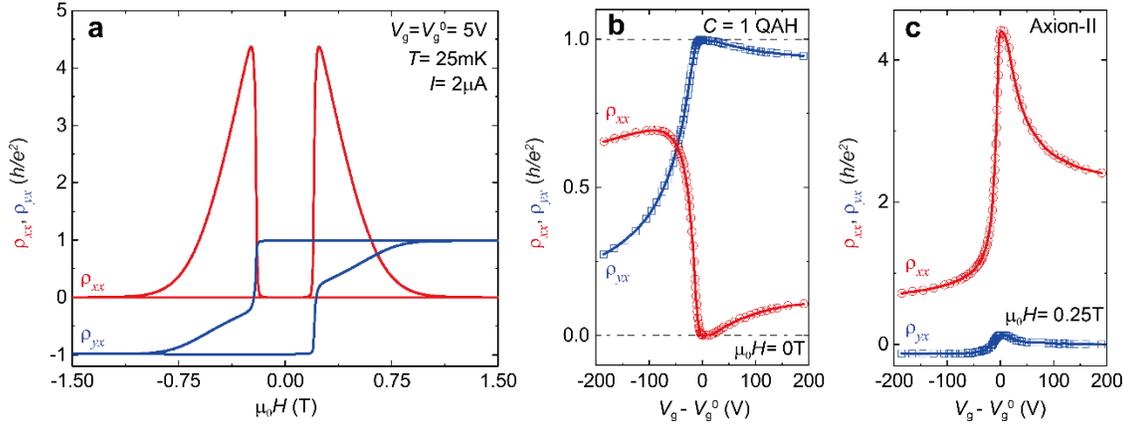

**Extended Data Fig. 1| QAH and axion insulator states in Sample S1. a,** $\mu_0H$-dependent $\rho_{xx}$ (red) and $\rho_{yx}$ (blue) measured at $V_g = V_g^0 = 5$ V and $T = 25$ mK. **b,** ($V_g$-$V_g^0$)-dependent $\rho_{xx}(0)$ (red) and $\rho_{yx}(0)$ (blue) for the $C = 1$ QAH state measured at $\mu_0H = 0$ T and $T = 25$ mK. **c,** ($V_g$-$V_g^0$)-dependent $\rho_{xx}(0.25T)$ (red) and $\rho_{yx}(0.25T)$ (blue) for the Axion-II state measured at $\mu_0H = 0.25$ T and $T = 25$ mK.

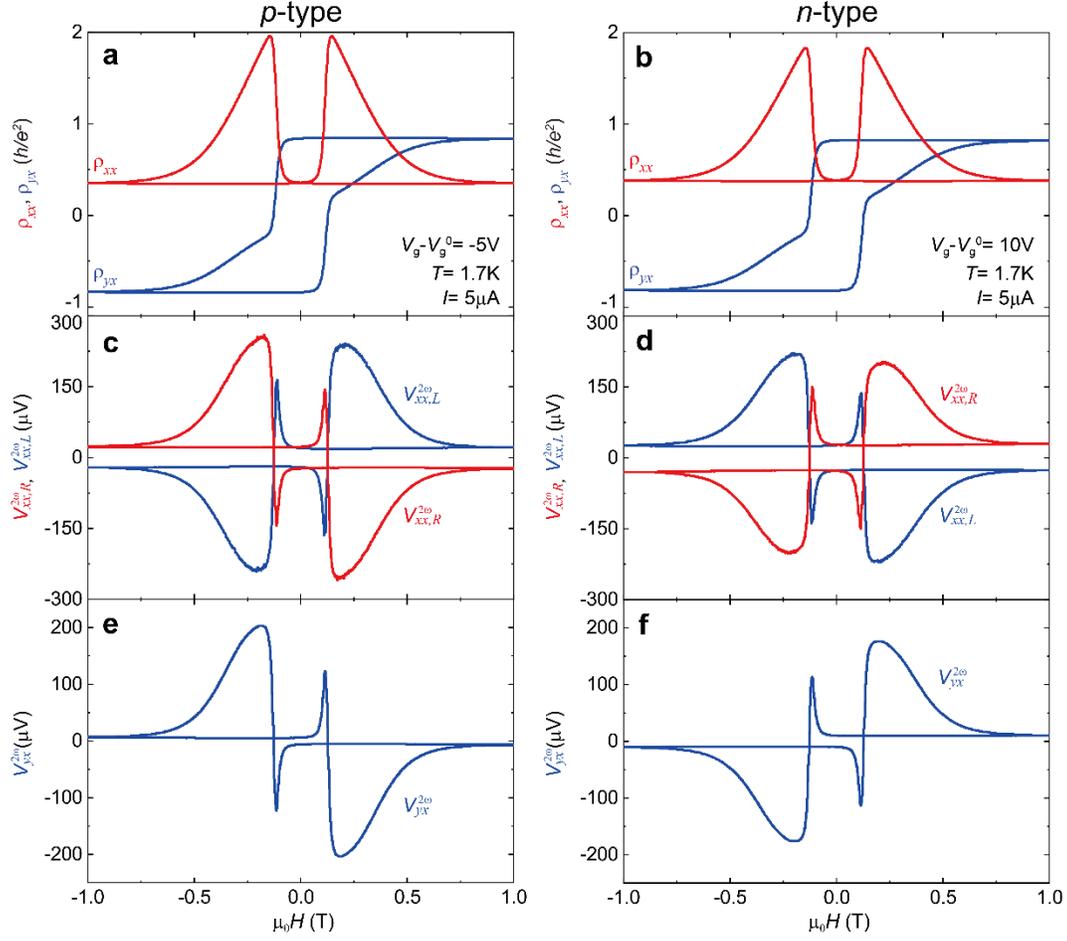

**Extended Data Fig. 2| Linear and nonlinear transport of Sample S1 measured at $T$ = 1.7 K. a, b,** $\mu_0H$-dependent $\rho_{xx}$ (red) and $\rho_{yx}$ (blue) measured at $(V_g - V_g^0)$ = -5 V (**a**) and 10 V (**b**). **c, d,** $\mu_0H$-dependent $V_{xx,R}^{2\omega}$ (red) and $V_{xx,L}^{2\omega}$ (blue) measured at $(V_g - V_g^0)$ = -5 V (**c**) and 10 V (**d**). **e, f,** $\mu_0H$-dependent $V_{yx}^{2\omega}$ measured at $(V_g - V_g^0)$ = -5 V (**e**) and 10 V (**f**).

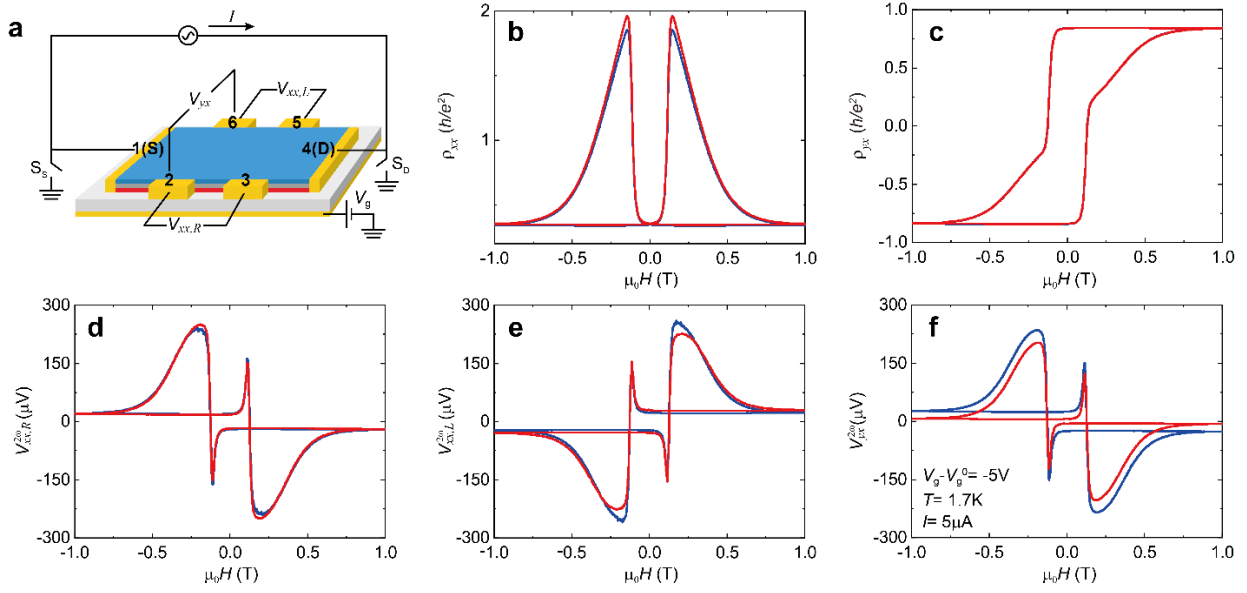

**Extended Data Fig. 3| Comparison of nonlinear transport measured using different ground electrodes in Sample S1. a,** Schematic of the magnetic TI sandwich Hall bar device. Note that in the measurements for electrodes 1 and 4, only one is grounded while the other remains floating. **b-f,** $\mu_0 H$-dependent $\rho_{xx}$ (**b**), $\rho_{yx}$ (**c**), $V_{xx,R}^{2\omega}$ (**d**), $V_{xx,L}^{2\omega}$ (**e**), and $V_{yx}^{2\omega}$ (**f**) measured at $T = 1.7$ K and ($V_g - V_g^0$) = -5 V. The red (blue) curve corresponds to measurements with electrode 1 (4) grounded.

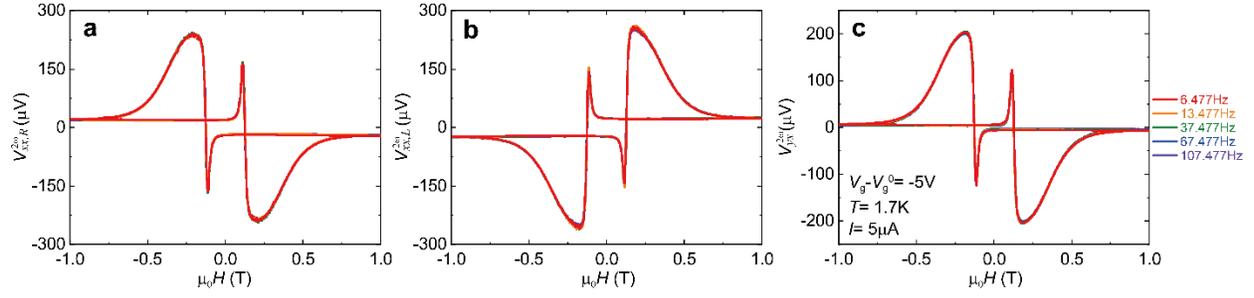

**Extended Data Fig. 4| Nonlinear transport measured across different frequencies in Sample S1. a-c,** $\mu_0 H$-dependent $V_{xx,R}^{2\omega}$ (**a**), $V_{xx,L}^{2\omega}$ (**b**), and $V_{yx}^{2\omega}$ (**c**) measured at $T$ = 1.7 K and $(V_g - V_g^0)$ = -5 V.

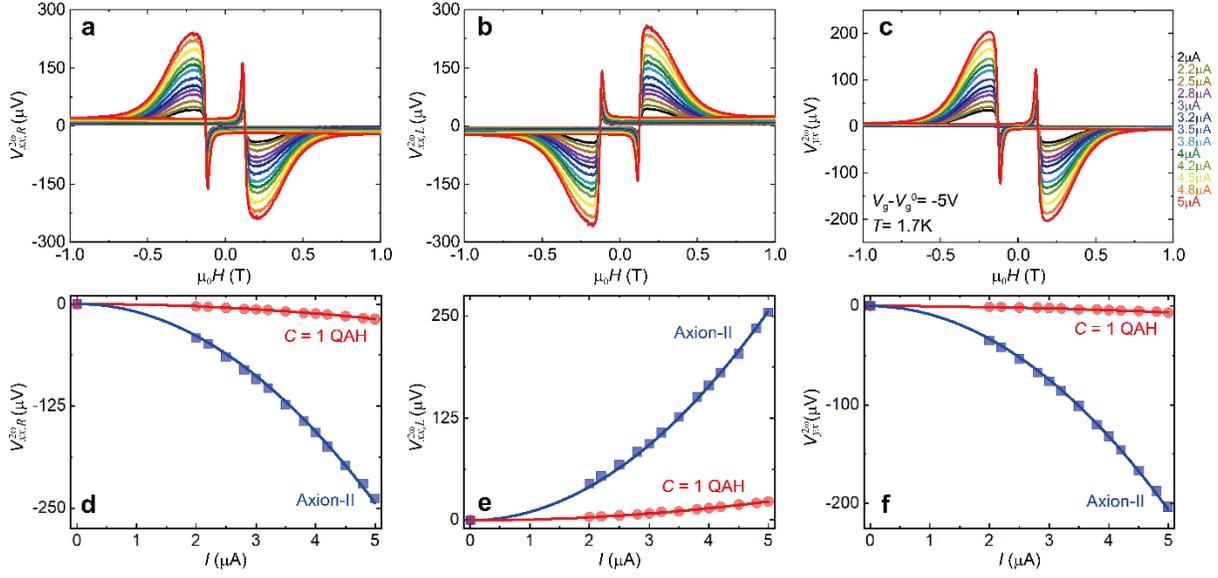

**Extended Data Fig. 5| Nonlinear transport measured across different currents in Sample S1.**
**a-c,** $\mu_0H$-dependent $V_{xx,R}^{2\omega}$ (**a**), $V_{xx,L}^{2\omega}$ (**b**), and $V_{yx}^{2\omega}$ (**c**) measured at $T = 1.7$ K and $(V_g - V_g^0) = -5$ V. **d-f**, Quadratic dependence of the nonlinear response $V_{xx,R}^{2\omega}$ (**d**), $V_{xx,L}^{2\omega}$ (**e**), and $V_{yx}^{2\omega}$ (**f**) on applied current $I$ for the $C = 1$ QAH (red circles) and Axion-II (blue squares) states. The blue and red curves are parabolic fits.

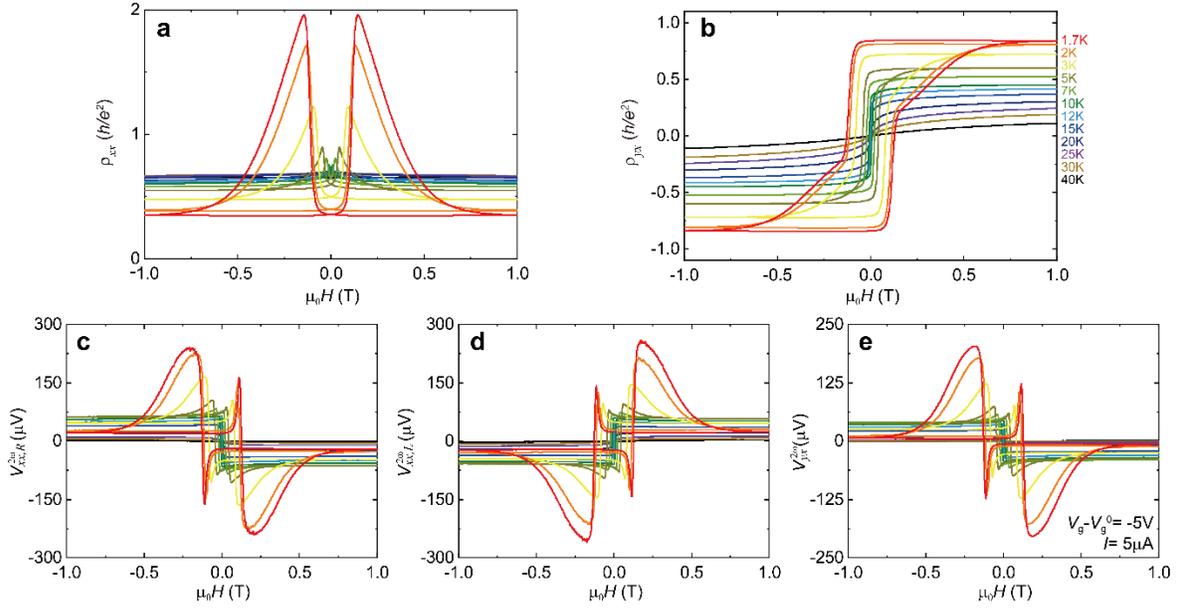

**Extended Data Fig. 6| Linear and nonlinear transport measured at different temperatures in Sample S1. a, b,** $\mu_0H$-dependent $\rho_{xx}$ (**a**) and $\rho_{yx}$ (**b**). The Curie temperature $T_c$ of Sample S1 is ~20 K. **c-e,** $\mu_0H$-dependent $V_{xx,R}^{2\omega}$ (**c**), $V_{xx,L}^{2\omega}$ (**d**), and $V_{yx}^{2\omega}$ (**e**). All measurements are performed at ($V_g - V_g^0$) = -5 V.

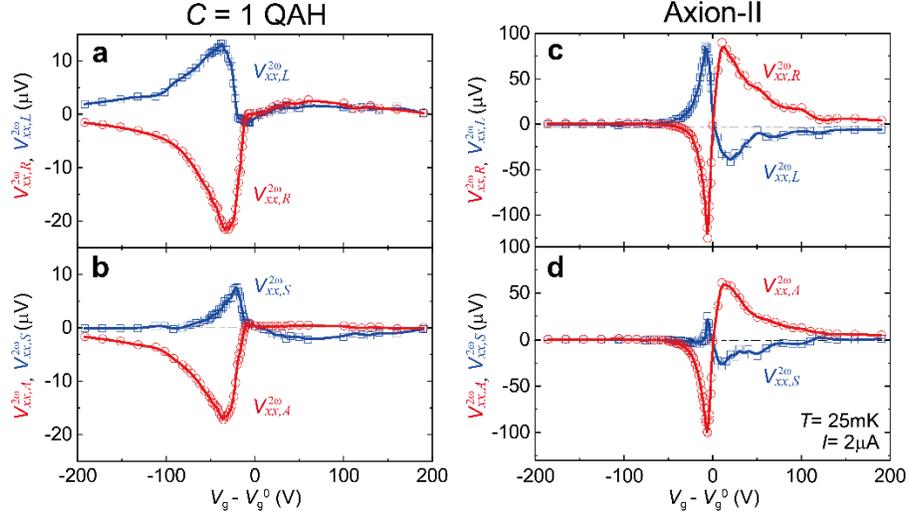

**Extended Data Fig. 7| Gate dependence of nonlinear longitudinal transport in Sample S1. a,** ($V_g$-$V_g^0$)-dependent $V_{xx,R}^{2\omega}$ (red) and $V_{xx,L}^{2\omega}$ (blue) for the $C$ = 1 QAH state. **b,** ($V_g$-$V_g^0$)-dependent $V_{xx,A}^{2\omega}$ (red) and $V_{xx,S}^{2\omega}$ (blue) for the $C$ = 1 QAH state. **c, d,** Same as (**a**) and (**b**), but for the Axion-II state. All measurements are performed at $T$ = 25 mK.

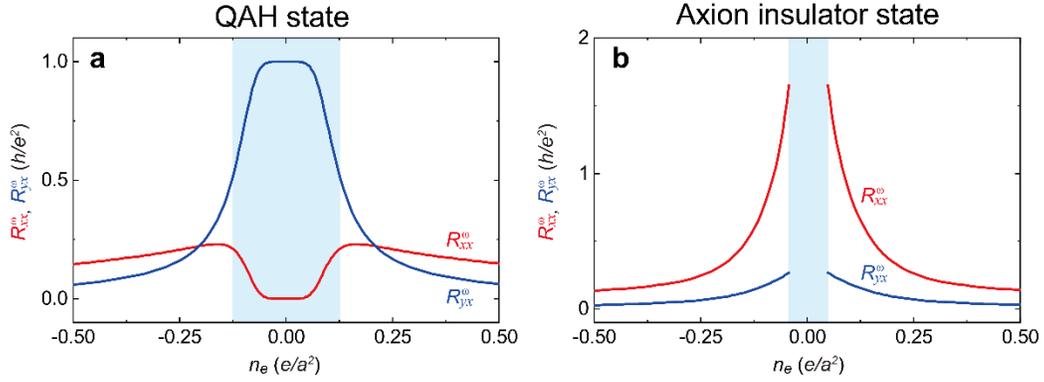

**Extended Data Fig. 8| Calculated first-order resistance as a function of carrier density for QAH and axion insulator states. a, b,** The calculated first-order resistance $R^{\omega}_{xx}$ (red) and $R^{\omega}_{yx}$ (blue) as a function of $n_e$ for the QAH (**a**) and axion insulator (**b**) states. The magnetic exchange gap regime is highlighted by the blue-shaded region, in which the quantized transport appears for the QAH state, but for the axion insulator state, the resistance will diverge in our calculations due to the vanishing of transmission coefficients.

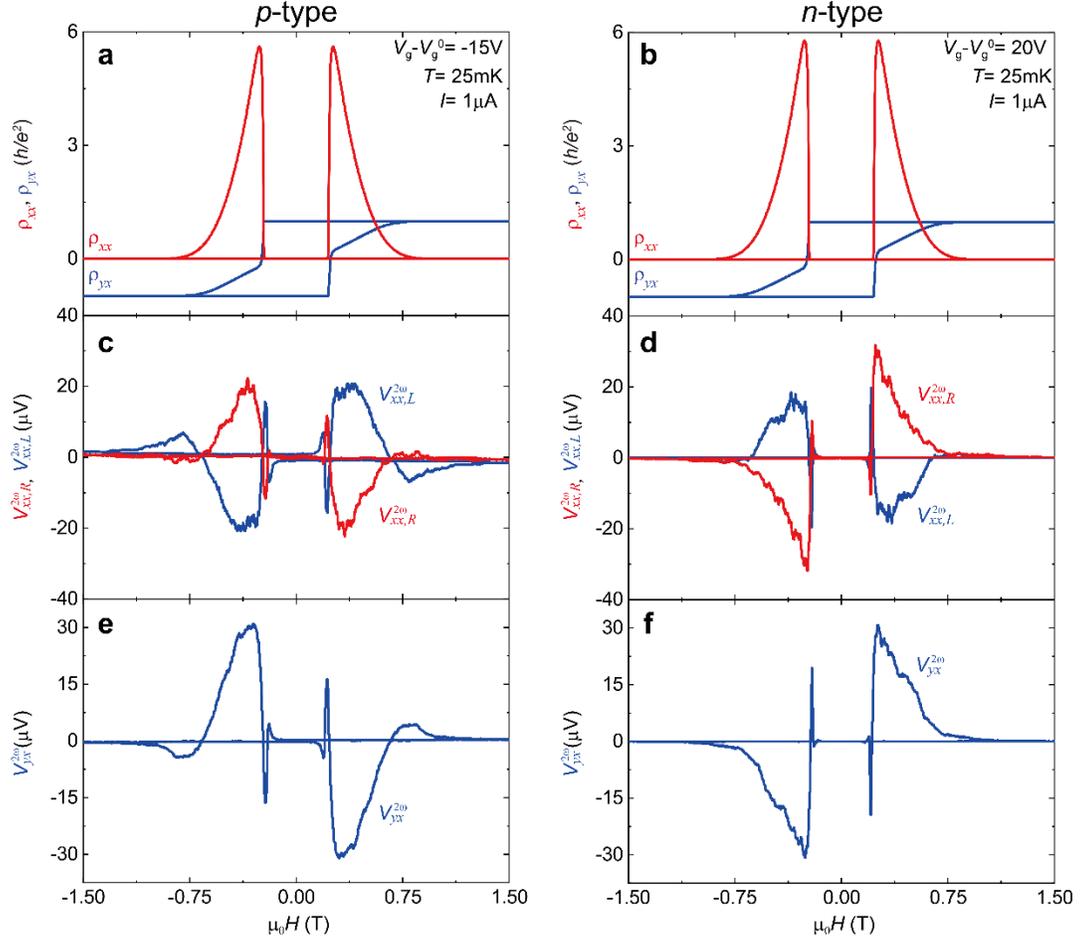

**Extended Data Fig. 9| Nonlinear transport in Sample S2 at $T$ = 25 mK. a, b,** $\mu_0H$-dependent $\rho_{xx}$ (red) and $\rho_{yx}$ (blue) measured at $(V_g - V_g^0)$ = -15 V (**a**) and 20 V (**b**). **c, d,** $\mu_0H$-dependent $V_{xx,R}^{2\omega}$ (red) and $V_{xx,L}^{2\omega}$ (blue) measured at $(V_g - V_g^0)$ = -15 V (**c**) and 20 V (**d**). **e, f,** $\mu_0H$-dependent $V_{yx}^{2\omega}$ (blue) measured at $(V_g - V_g^0)$ = -15 V (**e**) and 20 V (**f**).

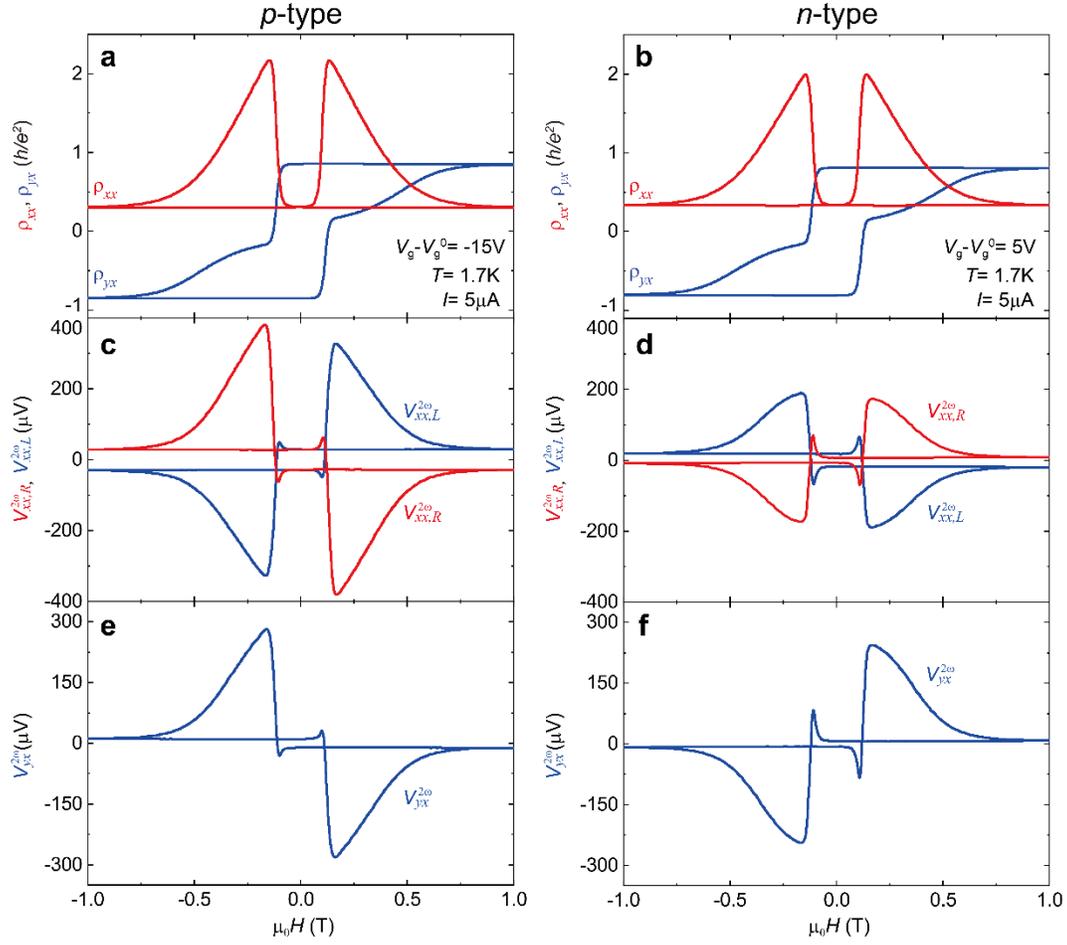

**Extended Data Fig. 10| Nonlinear transport in Sample S2 at $T$ = 1.7 K . a, b,** $\mu_0H$-dependent $\rho_{xx}$ (red) and $\rho_{yx}$ (blue) measured at $(V_g - V_g^0)$ = -15 V (**a**) and 5 V (**b**). **c, d,** $\mu_0H$-dependent $V_{xx,R}^{2\omega}$ (red) and $V_{xx,L}^{2\omega}$ (blue) measured at $(V_g - V_g^0)$ = -15 V (**c**) and 5 V (**d**). **e, f,** $\mu_0H$-dependent $V_{yx}^{2\omega}$ (blue) measured at $(V_g - V_g^0)$ = -15 V (**e**) and 5 V (**f**).